Latex file

-------------------------------------------------------------------------------

\documentstyle[12pt]{article}
\def\double{\baselineskip 24pt \lineskip 10pt}

\begin{document}
  \begin{titlepage}
  \begin{center}
\vspace{.8cm}
\Large
{\bf Oscillating Friedman Cosmology\\}
\vspace{.8cm}
\normalsize
\large{Mariusz P.~D\c{a}browski} \\
\normalsize
\vspace{.5cm}
{\em Institute of Physics, University of Szczecin, \\ Wielkopolska 15,
70-451 Szczecin, Poland}\footnote{E-mail:mpdabfz@uoo.univ.szczecin.pl}\\
\vspace{.3cm}
{\em and}\\
\vspace{.3cm}
{\em Astronomy Centre, University of Sussex, \\ Falmer,
Brighton BN1 9QH, U.~K.}\\
\end{center}
\vspace{.6cm}
\baselineskip=24pt
\begin{abstract}
\noindent
\double

The non-singular, oscillating Friedman cosmology within the framework of
General Relativity is considered. The general oscillatory solution given in
terms of elliptic functions and the conditions for its existence are discussed.
It is shown that the wall-like-matter and the small, but negative cosmological
constant are required for oscillations. The oscillations can , in principle,
be deep enough to allow standard hot universe processes like recombination and
nucleosynthesis.

It is shown that the wall-like-matter and string-like-matter can be
interpreted as scalar fields with some potentials. This may give another
candidate for the dark matter which may be compatible with observational data.
For an exact elementary oscillatory solution it is shown that the associated
scalar field potential is oscillating as well.

\end{abstract}
\vspace{.6cm}
{\small short title: oscillating cosmology}\\
\vspace{.6cm}

\end{titlepage}

\vspace{0.6cm}
\section{INTRODUCTION}
\vspace{0.6cm}
\double

The idea of a quasi-static and ever-lasting nonsingular universe seems still
to be very attractive among cosmologists. It was first Einstein \cite{E} who
did not believe his equations could really give in general non-static
solutions, until the clear arguments of Friedman \cite{F}. With the discovery
of
Hubble redshifting and the mathematical work of Hawking and Penrose \cite{HE}
about the inevitability of singularities rather few attention was paid
to non-singular models of the universe until the eighties. However,
singularity-free Friedman models usually called bouncing models were considered
early in thirties by Robertson \cite{ROB}. These models appear whenever the
cosmological constant, introduced by Einstein to balance gravitational
attraction, has a value between zero and a certain fixed number that gives
the Einstein Static Universe $ 0 < \Lambda < \Lambda_{st} $ \cite{RR}.
The universe begins with some minimum value of the scale factor
and then expands forever. Since the idea of inflation was established
\cite{G,L},
the situation changed because the scalar field responsible
for the scenario necessarily had to break the energy conditions giving the
effective negative pressure (repulsion) so much demanded by Einstein.

If we raise the question what are the ways to avoid singularities one can
distinguish a couple of approaches. In some opinions one should consider a
quantum gravity as an appropriate gravity theory at high densities. However,
most of the approaches refer to the alternative to General Relativity
classical gravity theories. Worth mentioning is the early proposal made by
Cartan \cite{C1,C2,C3} of nonsymmetric connection gravity theory with
torsion in which both isotropic and anisotropic universes start with a very
small but non-zero size \cite{TR,KOP}. Petry \cite{P1,P2} considered a
covariant gravity theory in which isotropic universes are
always bouncing \cite{P3}. In the nonsymmetric metric gravity theory of
Moffat \cite{M1,M2,M3} the black hole is replaced by a superdense object
\cite{M4}. Then, due to a repulsive force generated naturally in the
theory, non-singular cosmological solutions should also be a rule. Other
examples are the higher gravity theories such as: the second-order $R +
\epsilon R^2$ theory \cite{BO} with a bounce solution for
negative curvature \cite{P,CM}, the fourth-order
$R + \lambda R_{\mu \nu}R^{\mu \nu}/R + \tau R_{\alpha \beta \gamma \delta}
R^{\alpha \beta \gamma \delta}/R$ theory \cite{XE} and Brandenberger's
\cite{B} important proposal in which all isotropic solutions are the de
Sitter, though non-singular, type.

If we stay within General Relativity we can keep the simple Robertson-Walker
geometry and either assume a specific equation of state for Friedman models
\cite{IR1,IR2} or explore the scalar field coupled to gravity.
Both approaches are actually equivalent since the scalar field can mimic
different equations of state \cite{BA,LID}.

The aim of this paper is to consider singularity-free bouncing Friedman
universes within General Relativity admitting some fraction of negative
pressure matter usually believed to describe cosmic strins and domain walls
which we will be calling from now on, for the sake of some generality,
string-like-matter and wall-like-matter \cite{ST}. The physical motivation for
taking exotic matter into consideration is that we still do not have a solution
of the dark non-baryonic matter problem. One can strongly believe, in the
context of particle physics, that the exotic matter may be a very good
candidate for the dark matter. If it appeared that the exotic matter existed
one would believe in the non-singular universes of this paper. However, there
exixts some objections referring to the compatibility of the domain walls with
observations \cite{TUR,KAR}.
Although some other suggestions emerged \cite{GEL,GO}, it is believed that
even if walls formed in the early universe, they should decay by the present
\cite{RS,C}. In order to avoid the domain walls and perhaps to give more
freedom in compatibility with observations we replace the string-like-matter
and wall-like-matter by scalar fields. We postulate that scalar fields may be
the candidates for the dark matter in the present era of the universe
\cite{MCD,BA}. Our procedure is similar to the procedure applied for
inflationary universes \cite{BAR}. First we assume the exotic equation of state
and then we derive the form of the scalar field potentials.

Among non-singular solutions we select those which are oscillating i.e. those
for which the scale factor oscillates between certain fixed values. Such
solutions
first mentioned by Harrison \cite{HAR} and recently discussed by Kardashev
\cite{KAR}
originate from a qualitative change for bouncing models \cite{ROB} after
taking the negative pressure matter into account and admitting negative
cosmological constant. These solutions have both contracting and expanding
phase, so they may be compatible with the astronomical data. They last
infinitely in time which might be satisfactory at least from philosophical
point of view.

Oscillating models might also have another advantage. Since they are
quasi-static and require a similar balance to static models, then, according
to the analysis of Gibbons \cite{GIB1,GIB2}, they should posses very large
entropy
with value close a maximum admissible for the Einstein Static Universe.

The plan of the paper is as follows. In Section 2 we discuss the existence of
oscillating Friedman solutions and give the general exact oscillating solution
in terms of the elliptic functions which has not been studied before.
Also, we discuss a possibility to get very deep oscillations (i.e. those which
can
reach very high density to allow at least some standard hot universe processes)
and to reduce the amount of exotic matter in form of wall-like-matter. In
Section 3 we present the exact elementary non-oscillatory solutions which
complete the discussion of Section 2. In Section 3
we give an alternative scalar field interpretation for exotic fluids necessary
to force the universe to oscillate. We choose a very simple model of the scalar
fields of which potential and kinetic energy are proportional, following the
discussion of Barrow and Saich \cite{BS}. For a monotonic solution and the
oscillating Kardashev's solution we present exact scalar fields and their
potentials. It appears that for oscillating solutions for the scale factor
the potentials are also periodic in the scalar fields. In Section 5 we comment
on the results.

\vspace{.6cm}
\section{OSCILLATING UNIVERSES}
\vspace{.6cm}

Following our earlier discussion of exact analytic solutions of the Friedman
equation \cite{D86A,D89} we generalize it to the case
when other negative pressure fluids are present, in particular wall-like-matter
whose energy density scales as $ R^{-1} $ (R is the scale factor)
\cite{ST}.  The Friedman equation has the form
\begin{equation}
\left( \frac{dR}{d\tau} \right)^{2} = C_{r} + C_{m}R - k'R^2 + C_{w}R^3 +
\frac{\Lambda}{3}R^4   ,
\end{equation}
where $ \tau $ is the conformal time defined by the cosmic time as
\begin{equation}
d\tau = \frac{dt}{R}   ,
\end{equation}
and $k' = k - C_{s}$, k is the curvature index, and the constants
$C_{r}, C_{m}, C_{s}, C_{w}$ are constants responsible for the density of
radiation, nonrelativistic matter, string-like-matter (whose energy density
scales as $ R^{-2} $) and wall-like-matter respectively, and
$\Lambda$ is the cosmological constant. The Fried\-man equation (2.1)
in terms of the so-called reduced blackbody temperature \footnote{The reduced
blackbody temperature $T(\tau)$ is proportional to the temperature of the
microwave background if $\alpha \neq 0$, and it is only interpreted as an
inverse function of $R(\tau)$ if $\alpha = 0$ \cite{CG}}
\begin{equation}
T = \Lambda_{c}^{ - \frac{1}{2}} R^{-1},
\end{equation}
where
\begin{equation}
\Lambda_{c}^{ - \frac{1}{2}} \equiv \frac{3}{2} C_{m} ,
\end{equation}
is \cite{CG,D89}
\begin{equation}
\left( \frac{dT}{d\tau} \right)^{2} = \alpha T^4 + \frac{2}{3} T^3 - k' T^2 +
\beta T + \frac{\lambda}{3}   ,
\end{equation}
where the dimensionless parameter
\begin{equation}
\beta = C_{w}\Lambda_{c}^{ - \frac{1}{2}}
\end{equation}
is responsible for the density of wall-like-matter ($\alpha = C_{r}\Lambda_{c},
\lambda = \Lambda / \Lambda_{c}$).
In order to make qualitative analysis of the solutions of (2.5) we use the
method of the associated mechanical system \cite{CG,CHA}. In fact (2.5) can be
considered as the energy equation of a one-dimensional mechanical system with
$T$ as a coordinate, the
kinetic energy $\left( dT / d\tau \right)^2 \geq 0$, the potential
\begin{equation}
V_{k',\alpha,\beta}(T) \equiv - \alpha T^4 - \frac{2}{3} T^3 + k' T^2 - \beta T
 =
- Q_{k',\alpha,\beta,\lambda}(T) + \frac{\lambda}{3}  ,
\end{equation}
where
\begin{equation}
Q_{k',\alpha,\beta,\lambda}(T) \equiv \alpha T^4 + \frac{2}{3} T^3 - k' T^2 +
\beta T + \frac{\lambda}{3}  ,
\end{equation}
and the total energy $\frac{\lambda}{3}$.

The general shape of the potential (2.7) depends on the values of the factors
$k', \alpha$ and $\beta$. Instead of the full discussion of the solutions (2.5)
we will concentrate on
special cases which allow the universe to oscillate in time between the two
fixed values of the scale factor $R(\tau)$ (or $T(\tau)$) \cite{KAR}. In order
to have oscillations at least two kinds of
matter are necessary i.e. the cosmological constant and the wall-like-matter.
To make our considerations easier we put $\alpha = 0$ (no radiation)
\footnote{ One can consider the full formula
(2.7) that includes radiation as well, and the roots of the cubic equation
$ - \alpha T^3 - \frac{2}{3} T^2 + k' T - \beta $ which we will call $T_{1},
T_{2}$ and $T_{3}$ have to fulfil the conditions
\begin{eqnarray}
T_{1} + T_{2} + T_{3} & = & - \frac{2}{3\alpha}  \nonumber ,\\
T_{1}T_{2} + T_{1}T_{3} + T_{2}T_{3} & = & - \frac{k'}{\alpha}  \nonumber ,\\
T_{1}T_{2}T_{3} & = & - \frac{\beta}{\alpha}  \nonumber  .
\end{eqnarray}
It follows from these relations that there should be at least one negative real
root and the two real or complex conjugate roots, which means that radiation
pressure can only change the left branch of the curve of the potential
$V_{\alpha,k',\beta}(T)$ (i.e. negative $T(\tau)$ - cf. Figs.1-3) which is not
physically relevant.}

The potential (2.7) is
then
\begin{equation}
V_{k',\beta} = - \frac{2}{3} T^3 + k' T^2 - \beta T  ,
\end{equation}
and it has one double root $T_{1,2} = \frac{3}{4}k'$ for $\beta = \frac{3}{8}
k'^2$ and two real roots for $\beta < \frac{3}{8}k'^2$ as well as the root at
$T = 0$. Its corresponding extrema are a double one at $\tilde{T}_{1,2} =
 \frac{1}{2}k'$ for $\beta = \frac{1}{2} k'^2$ (which is an inflection point as
well) and two real ones for $\beta < \frac{1}{2}k'^2$.

The general solution of (2.5) given in terms of the We\-ier\-strass
el\-lip\-tic
${\cal P}$ function is
\begin{equation}
T(\tau) = \frac{3\sqrt{\alpha} {\cal P}'(\tau) - {\cal P}(\tau) - \frac{k'}{12}
- \frac{1}{48} \alpha \beta k'^2}{6 \alpha {\cal P}(\tau) - \frac{1}{6} -
k' \alpha}     ,
\end{equation}
where ${\cal P}$ is defined by the equation
\begin{eqnarray}
{\cal P}'(\tau) \equiv \frac{d{\cal P}}{d\tau} = \sqrt{4{\cal P}^3 -
g_{2}{\cal P} - g_{3}}   \nonumber   ,
\end{eqnarray}
and the invariants
\begin{eqnarray}
g_{2} & = & \frac{k'^2}{12} + \frac{\alpha\lambda}{3} - \frac{\beta}{6}  ,\\
g_{3} & = & 6^{-3} \left( k'^3 - 2\lambda - 3k'\beta \right) -
\frac{\alpha}{2} \left( \frac{k'\lambda}{9} + \frac{\beta^2}{8} \right)  .
\end{eqnarray}
For $\alpha = 0$ case (no radiation) corresponding to the potential (2.9) we
have the general solution of (2.5) in the elliptic form as well
\begin{equation}
T(\tau) = 6 \left[ {\cal P}(\tau) + \frac{k'}{12} \right]   ,
\end{equation}
with the invariants given by (2.11)-(2.12) for $\alpha = 0$ and the
discriminant
\cite{D86A}
\begin{equation}
\Delta_{k',\beta,\lambda} \equiv g_{2}^3 - 27 g_{3}^2 = 2^{-4}3^{-3} \left[
\beta^2 \left( \frac{3}{4}k'^2 - 2\beta \right) + k'\lambda \left( k'^2 -
3\beta \right) - \lambda^2 \right]  ,
\end{equation}
i.e.
\begin{equation}
\Delta_{k',\beta,\lambda} = - 2^{-4}3^{-3} \left( \lambda - \lambda_{+} \right)
\left( \lambda - \lambda_{-} \right)   ,
\end{equation}
where the critical values of $\lambda$ are
\begin{equation}
\lambda_{\pm} = \frac{1}{2}k' \left( k'^2 - 3\beta \right) \pm \frac{1}{2}
\sqrt{k'^2 \left( k'^2 - 3\beta \right)^2 + 4\beta^2 \left( \frac{3}{4}k'^2
- 2\beta \right)}   .
\end{equation}
The associated mechanical system method is as follows. Having the exact shape
of the potential (2.7) (Figs.1-3) we cut the curves along the constant
$\lambda$ lines. The point associated with the universe can only move in the
upper part of the plane and we look for the possible solutions.
We consider only the particular cases given by (2.13) where oscillations are
possible ($T_{1,2}$ zeros and $\tilde{T}_{1,2}$ extrema of (2.9)). These are:

A. If $k' > 0$ and $\frac{\beta}{k'^2} < \frac{3}{8}$ we have
\begin{eqnarray}
T_{1,2} & = & \frac{3}{4} \left( k' \mp \frac{1}{3} \sqrt{9k'^2 - 24\beta}
\right) \nonumber ,\\
\tilde{T}_{1,2} & = & \frac{1}{2} \left( k' \mp \sqrt{k'^2 - 2 \beta} \right)
\nonumber   ,
\end{eqnarray}
and the oscillations corresponding to a constant $\lambda$ lines for
\begin{eqnarray}
\lambda_{-} < \lambda < 0   \nonumber
\end{eqnarray}
with $\lambda_{-}$ given by (2.16) are possible in the central well of the
potential (2.9) (Fig.1).

B. If $k' > 0$ and $\frac{\beta}{k'^2} = \frac{3}{8}$ we have
\begin{eqnarray}
T_{1} & = & T_{2} = \tilde{T}_{2} = \frac{3}{4}k' \nonumber ,\\
\tilde{T}_{1} & = & \frac{k'}{4} \nonumber   ,
\end{eqnarray}
and the oscillations are possible for
\begin{eqnarray}
- \frac{1}{8}k'^3 = \lambda_{-} < \lambda < \lambda_{+} = 0 \nonumber   ,
\end{eqnarray}
i.e. in the central well of the appropriate potential (2.9) (Fig.2).

C. If $k' > 0$ and $\frac{3}{8} < \frac{\beta}{k'^2} < \frac{1}{2}$ we have
\begin{eqnarray}
T_{1} & = & T_{2} = 0   \nonumber  ,\\
\tilde{T}_{1,2} & = & \frac{1}{2} \left( k' \mp \sqrt{k'^2 - 2 \beta} \right)
\nonumber   ,
\end{eqnarray}
and the oscillations are possible for
\begin{eqnarray}
\lambda_{-} < \lambda < \lambda_{+} < 0   \nonumber   ,
\end{eqnarray}
with $\lambda_{\pm}$ given by (2.15) (Fig.3).

There is apparently a double extremum for $T = \frac{k'}{2}$ if
$\frac{\beta}{k'^2} = \frac{1}{2}$, but it is also an inflection point of
(2.9) and this case does not allow any oscillating solutions at all.

In all considered cases $A, B, C$ the discriminant (2.15) is positive and the
fundamental periodicity cell is a rectangle (Fig.4) \cite{TRI}. The roots
$e_{1}, e_{2}, e_{3}$ of the equation
\begin{equation}
4y^3 + g_{2}y + g_{3} = 0
\end{equation}
are all real and since $e_{1} + e_{2} + e_{3} = 0$, at least one of them must
be negative. One of the elementary periods $\omega$ is pure real and the other
$\omega^{'}$ is pure imaginary. The roots are given by the relations
\begin{eqnarray}
{\cal P}(\omega) = e_{1}  \nonumber  ,\\
{\cal P}(\omega + \omega^{'}) = e_{2}  \nonumber  ,\\
{\cal P}(\omega^{'}) = e_{3}  \nonumber  ,
\end{eqnarray}
and
\begin{equation}
e_{1} < e_{2} < e_{3}    .
\end{equation}
However, since we consider the solutions of the equation for $T(\tau)$ (cf.
(2.5)) then we should define the roots of the equation
\begin{equation}
\frac{2}{3} T^3 - k' T^2 + \beta T + \frac{\lambda}{3} = 0   ,
\end{equation}
which are
\begin{eqnarray}
T_{min} = 6e_{3} + \frac{k'}{2}  \nonumber   ,\\
T_{max} = 6e_{2} + \frac{k'}{2}  \nonumber   ,\\
T_{recol} = 6e_{1} + \frac{k'}{2}  \nonumber   ,
\end{eqnarray}
and
\begin{equation}
T_{min} < T_{max} < T_{recol}   ,
\end{equation}
where $T_{recol}$ refers to a minimum of $T(\tau)$, i.e. a maximum of $R(\tau)$
for the recollapsing model associated with (2.13) and $T_{min}, T_{max}$ refer
to a minimum and a maximum of $T(\tau)$ for the oscillating model associated
with (2.13). From (2.19) we conclude that $T_{min}, T_{max}$ and $T_{recol}$
must be real and positive since $T_{min} + T_{max} + T_{recol} = \frac{3}{2}$
and $T_{min}T_{max}T_{recol} = - \lambda/2$ and $\lambda$ is
negative in the cases $A, B, C$.

The general oscillatory solution for all the cases $A, B, C$ can be expressed
in terms of the Weierstrass $\zeta$ function \cite{CG}
\begin{equation}
\frac{1}{T(\tau)} = \Lambda_{c}^{\frac{1}{2}} = \frac{1}{T_{min}} +
\sqrt{\frac{3}{\lambda}} \left[ \zeta(\tau - \tau_{d}) - \zeta(\tau + \tau_{d})
+ 2\zeta(\tau_{d}) \right]   ,
\end{equation}
and the expression for the cosmic time, from (2.2), is
\begin{equation}
\left( \frac{\Lambda}{3} \right)^{\frac{1}{2}} t(\tau) =
\tau \left[ \left( \frac{\lambda}{3} \right)^{\frac{1}{2}} \frac{1}{T_{min}} +
2 \zeta(\tau_{d}) \right] + \ln \frac{\sigma(\tau_{d} - \tau)}
{\sigma(\tau_{d} + \tau)}   ,
\end{equation}
so for $\tau = 0$ $t(\tau) = 0$ and $T = T_{min}$ (Fig.6). In both formulas
(2.21) and (2.22) the conformal time is real but the zeros of the function
$T(\tau)$, namely $ - \tau_{d}$ and $\tau_{d}$ are imaginary. Also, $\lambda$
in
these expressions should be taken negative according to the general results of
the existance of the oscillatory solutions and then $\sqrt{3/\lambda}$
and $\sqrt{\Lambda/3}$ are imaginary.
The periods are given by \cite{TRI}
\begin{eqnarray}
\omega = \int_{T_{min}}^{T_{max}} \frac{dT}{\sqrt{\frac{2}{3}(T - T_{min})
(T - T_{max})(T - T_{recol})}}   \nonumber  ,\\
\omega^{'} = \int_{T_{max}}^{T_{recol}} \frac{dT}{\sqrt{\frac{2}{3}(T -
T_{min})
(T - T_{max})(T - T_{recol})}}   \nonumber  ,
\end{eqnarray}
so $\omega$ is pure real and $\omega^{'}$ is pure imaginary.

Let us now consider another type of the exact oscillating universes, the ones
in which there is no dust (i.e. $C_{m} = 0$ in (2.1) and $\alpha = 0$). Because
of the definition (2.4) we cannot now use the relation (2.3) to work out the
limit $C_{m} \rightarrow 0$ from (2.5). Instead we use the method by putting
\begin{equation}
M(\tau) = \frac{1}{R(\tau)}   ,
\end{equation}
so (2.1) becomes
\begin{equation}
\left( \frac{dM}{d\tau} \right)^{2} = - k' M^2 + C_{w} M + \frac{\Lambda}{3}
{}.
\end{equation}
The potential analogous to (2.7) now is (Fig.7)
\begin{equation}
V_{k',C_{w}} = k' M^2 - C_{w} M   ,
\end{equation}
and provided $k' > 0$ we have oscillations for $\Lambda < 0$. In this case the
solution is elementary and oscillating \cite{KAR}. After a
deparametrization based on (2.2) the exact solution can be written down as
\begin{equation}
R(t) = - \frac{3}{2\Lambda} \left[ A \sin{t \sqrt{ - \frac{\Lambda}{3}}} +
C_{w} \right]  ,
\end {equation}
where
\begin{eqnarray}
C_{w} & > & A \equiv \sqrt{C_{w}^2 + \frac{4}{3} \Lambda k'}  ,\\
C_{w}^2 & > & - \frac{4}{3} \Lambda k' \nonumber ,\\
\Lambda & < & 0 \nonumber  ,
\end{eqnarray}
so the universe oscillates between $R_{min} = - (3/2)\Lambda(-A + C_{w})$ and
\\
$R_{max} = - (3/2)\Lambda(A + C_{w})$.

{}From the shape of the associated potential (2.7) for $\beta =
\frac{3}{8}k'^{2}$
(Case B, Fig.2) one can conclude that for \footnote{In fact the restrictions
given below are actually the strongest ones since the minimum size of the
oscillating universes in cases A, B, C is always bigger (i.e. $\tilde{T_{2}} <
\frac{3}{4}k'$ cf. Eq.(3.2)).}
\begin{eqnarray}
\lambda_{-} = - \frac{1}{8}k'^3 < \lambda < 0 \nonumber
\end{eqnarray}
the universe will oscillate between the fixed extrema of $R(\tau)$ and
the smallest admissible value of the scale factor will be slightly larger than
(cf. Section 3)
\begin{equation}
R_{u}(\tau) = \frac{1}{\Lambda_{c}^{\frac{1}{2}}\tilde{T}_{2}} =
2\frac{C_{m}}{k'}    ,
\end{equation}
so the mass density of nonrelativistic matter will be \cite{D86A}
\begin{equation}
\varrho_{m} = \frac{3c^2}{8\pi G} C_{m}R_{u}^{-3} = \frac{k'^3}{C_{m}^2}
\frac{3c^2}{64\pi G}   .
\end{equation}
If we assume $k' = k - C_{s} \approx 1$ (only $k = +1$ is possible here, since
$k' > 0$) which means that the amount of string-like-matter is small and most
of the exotic matter is in the form of wall-like-matter , then from (2.28)
\begin{eqnarray}
\varrho_{m} \sim C_{m}^{-2} \cdot 2 \cdot 10^{26} \frac{g}{cm}   \nonumber   .
\end{eqnarray}
In order to calculate $C_{m}$ (with the dimension [cm]) we should apply the
conservation law (2.28). For the maximum density at the nucleosynthesis scale
\begin{eqnarray}
\varrho_{m} \sim 10^{4} \frac{g}{cm^3}   \nonumber
\end{eqnarray}
we need
\begin{eqnarray}
C_{m} \sim 1.4 \cdot 10^{11} cm   \nonumber   ,
\end{eqnarray}
which can be achieved for instance either for the present nonrelativistic
matter mass density $\varrho_{m0} \sim 1.1 \cdot 10^{-45} \frac{g}{cm^3}$
(rather tiny) and the present radius $R_{0} \sim 5 \cdot 10^9$ light years
or for the present mass density $\varrho_{m0} \sim 1.1 \cdot 10^{-35}
\frac{g}{cm^3}$ (larger) and the smaller present radius $R_{0} \sim 5 \cdot
10^6$ light years. However, for the maximum density at recombination scale
\begin{eqnarray}
\varrho_{m} \sim 10^{-18} \frac{g}{cm^3}   \nonumber
\end{eqnarray}
we need
\begin{eqnarray}
C_{m} \sim 1.4 \cdot 10^{22} cm   \nonumber   ,
\end{eqnarray}
and it requires for example the present nonrelativistic matter density
$\varrho_{m0} \sim 1.1 \cdot 10^{-34} \frac{g}{cm^3}$ and the present radius
$R_{0} \sim 5 \cdot 10^9$ light years which is quite realistic. This means
that oscillations might allow at least some standard early universe processes
like nucleosynthesis or recombination.

The amount of wall-like-matter can be reduced if we lessen the value of
$k' = k - C_{s}$ (i.e. decrease the appropriate value of $C_{m}$ - cf.(2.28)),
since the density of walls is proportional to $k'$
\begin{eqnarray}
\beta = \frac{3}{8} k'^2 = \frac{3}{8} \left( k - C_{s} \right)^2  \nonumber
,
\end{eqnarray}
and the wall-like-matter is replaced by string-like-matter ( $0 < C_{s} < 1$ ).
However, in such a case both the maximum and minimum of the potential
(2.7) are approaching zero - the well becomes smaller which means that the
appropriate periods of oscillations are smaller.

Since the amplitude of oscillations corresponds to the length of the $\lambda
=$
const. lines contained inside the well of the potential (2.7) (Figs.1-3) one
can easily notice a very interesting feature of the models A and B. It is
that the closer to zero (necessarily negative) value of the cosmological
constant $\lambda$, the bigger the maximum, and the smaller the minimum size of
the universe can be, which means the longer period of oscillations. On the
other
hand, since we do not experience small period of oscillations (they can appear
for large negative $\lambda$ - Figs.1-3) we could obtain some restrictions on
the possible value of the cosmological constant in these models.

Finally, it is very interesting to mention that if we admit some fractional
equations of state
\begin{eqnarray}
p = \left( \varepsilon - 1 \right) \rho \nonumber
\end{eqnarray}
with $\varepsilon = \frac{1}{3} n $ (n is non-integer)
to derive the new Friedman equation different from (2.1), then a larger
variety of possibilities for oscillations appear \cite{HAR,D86B}.
It can be explained in terms of the associated mechanical system simply
because more wells of the potential are possible in which the universe can
oscillate. Such fractional equations of state are possible since the exotic
matter
may have an effective equation of state anywhere in a range between well
established
values. For instance for strings $\frac{2}{3} \leq \varepsilon_{s} \leq
\frac{4}{3}$ and
for walls $\frac{1}{3} \leq \varepsilon_{w} \leq \frac{4}{3}$ depending on
their
velocities \cite{KT} \footnote{It should be pointed out that string
or wall stress is in fact anisotropic and might be considered by using an
anisotropic model of spacetime rather than the isotropic one \cite{BGT}.
However, it seems that there is no reason for strings or walls to be
formed anisotropically throughout the whole universe and we can suitably
average the network over all directions \cite{HIN}}.

\vspace{.6cm}
\section{NON-OSCILLATING EXACT UNIVERSES}
\vspace{.6cm}
\setcounter{equation}{0}

In this Section we concentrate on some particular solutions given by elementary
functions. None of them is oscillatory but they complete the discussion of the
solutions from the cases $A, B, C$ as far as they can give some insight into
the
nature of the oscillatory solutions as well.

Generally, if the discriminant (2.14) vanishes the solution is elementary. From
(2.15) one can realize that it happens for $\lambda = \lambda_{+}$ and
$\lambda = \lambda_{-}$ with $\lambda_{\pm}$ given by (2.16). With the
condition of the vanishing discriminant the equation (2.5) factorizes to the
form
\begin{equation}
\left( \frac{dT}{d\tau} \right)^2 = \left( T - \tilde{T}_{1,2} \right)^2 \left(
\frac{2}{3}T + A \right)
\end{equation}
with
\begin{equation}
T = \tilde{T}_{1,2} = \frac{1}{2} \left( k' \mp \sqrt{k'^2 - 2\beta} \right)
\end{equation}
with $A =$ const.
With $\tilde{T}_{1,2}$ given by (3.2) we have two Einstein Static Universes in
the cases $A$ and $C$ (Figs.1 and 3). One of them should be stable because it
lies at the bottom of the well of the potential (2.7). In the case B i.e. for
$\frac{\beta}{k'^2} = \frac{3}{8}$ (3.2) reduces to
\begin{equation}
T = \tilde{T}_{1,2} = \frac{k'}{2} \left( 1 \mp \frac{1}{2} \right)   .
\end{equation}
With every stable solution for $\tilde{T}_{1}$ it is also associated the
elementary recollapsing solution for $\lambda = \lambda_{-}$. On the other hand
with each unstable solution for $\tilde{T}_{2}$ there are also associated two
asymptotic solutions for $\lambda = \lambda_{-}$.

Since the types of the elementary solutions of (3.1) in the cases $A$ and $C$
are analogous to the solutions in the case $B$ with only numerical difference
between (3.2) and (3.3) we will be discussing the latter case which is
mathematically simpler.
The stable Einstein Static Universe appears for $\lambda = \lambda_{-} =
- \frac{1}{8} k'^3$ and
\begin{equation}
R_{s} = \frac{1}{\Lambda_{c}^{\frac{1}{2}}\tilde{T_{1}}} = 6 \frac{C_{m}}{k'}
{}.
\end{equation}
The equation (3.1) for $\lambda = \lambda_{-}$ reads as
\begin{equation}
\left( \frac{dT}{d\tau} \right)^2 = \frac{2}{3} \left( T - \frac{k'}{4}
\right)^
2 \left( T - k' \right) ,
\end{equation}
and its solution for $T > k'$ describes the evolution of the closed universe
from the big-bang to the big-crunch in the form
\begin{equation}
R(\tau) = \frac{1}{\Lambda_{c}^{\frac{1}{2}}T(\tau)} =
\frac{\Lambda_{c}^{ - \frac{1}{2}}}{k' \left[ 1 +
 \frac{3}{4}\tan^2{\frac{1}{2}\sqrt{\frac{k'}{2}}\tau}
\right]}    .
\end{equation}
The unstable Einstein Static Universe appears for $\lambda = \lambda_{+} = 0$
and
\begin{equation}
R_{u} = \frac{1}{\Lambda_{c}^{\frac{1}{2}}\tilde{T_{2}}} = 2 \frac{C_{m}}{k'}
{}.
\end{equation}
The equation (3.1) for $\lambda = \lambda_{+} = 0$ reads as
\begin{equation}
\left( \frac{dT}{d\tau} \right)^2 = \frac{2}{3}T \left( T - \frac{3}{4}k'
\right)^2  .
\end{equation}
There are two asymptotic to (3.7) solutions of (3.8).
For $0 \leq T \leq \frac{3}{4}k'$ we have
\begin{equation}
R(\tau) = \frac{1}{\Lambda_{c}^{\frac{1}{2}} T(\tau)} = \frac{4}{3k'}
\Lambda_{c}^{ - \frac{1}{2}} \coth^2{\frac{1}{2}\sqrt{\frac{k'}{2}}\tau}  ,
\end{equation}
and according to (2.2)
\begin{equation}
t(\tau) = \frac{8\sqrt{2}}{3}\left( k'^3\Lambda_{c} \right)^{ - \frac{1}{2}}
\left[
\frac{1}{2}\sqrt{\frac{k'}{2}}\tau - \coth{\frac{1}{2}\sqrt{\frac{k'}{2}}\tau}
\right]  .
\end{equation}
Then for $T > \frac{3}{4}k'$ we have
\begin{equation}
R(\tau) = \frac{1}{\Lambda_{c}^{\frac{1}{2}} T(\tau)} =
\frac{4}{3k'}\Lambda_{c}^{ - \frac{1}{2}}
\tanh^2{\frac{1}{2}\sqrt{\frac{k'}{2}}\tau}  ,
\end{equation}
and
\begin{equation}
t( \tau) = \frac{ 8 \sqrt{2}}{3} \left( k'^3 \Lambda_{c} \right)^{ -
\frac{1}{2}} \left[ \frac{1}{2} \sqrt{ \frac{k'}{2}} \tau - \tanh{ \frac{1}{2}
\sqrt{ \frac{k'}{2}} \tau} \right]  .
\end{equation}

\vspace{.6cm}
\section{SCALAR FIELD COSMOLOGIES}
\vspace{.6cm}
\setcounter{equation}{0}

In this section we try to give an alternative interpretation for exotic matter
appearing in the Friedman equation (2.1) in terms of the scalar fields and
appropriate potentials. The motivation is that we still are looking for the new
candidates for the dark matter in cosmology and we may think about scalar
fields existing at the present era of the evolution as good candidates. Such
an interpretation seems to give more freedom
in imposing the observational constraints than string-like-matter and
wall-like-matter \cite{MCD}. Our procedure is a reverse of the standard
procedure dealing with scalar fields. Instead of starting with an explicit
potential to derive the equation of state, we start with the equations of state
for string-like-matter and wall-like-matter (cf. Section 2, formula (2.1)) in
order to derive the exact potentials \cite{BAR}. The energy density and
pressure
for two mutually noninteracting scalar fields minimally coupled to gravity can
be written as
\begin{eqnarray}
\rho_{i} = \frac{1}{2}\dot{\varphi}_{i}^2 + V_{i}(\varphi_{i})  ,\\
p_{i} = \frac{1}{2}\dot{\varphi}_{i}^2 - V_{i}(\varphi_{i})  ,
\end{eqnarray}
where i = 1,2. If, apart from this, we admit a barotropic fluid with the
equation of
state $p = (\nu - 1)\rho$ we will obtain the rewritten Friedman equation (2.1)
in the form \cite{WE}
\begin{equation}
3 \left( \frac{\dot{R}}{R} \right)^2 = 3H^2 = \rho +
\frac{1}{2}\dot{\varphi}_{1}^2
+ \frac{1}{2}\dot{\varphi}_{2}^2 + V_{1}(\varphi_{1}) + V_{2}(\varphi_{2})
- \frac{k}{R^2}  ,
\end{equation}
and the wave equations for both fields
\begin{eqnarray}
\ddot{\varphi}_{1} + 3H\dot{\varphi}_{1} + V_{1}'(\varphi_{1}) = 0  ,\\
\ddot{\varphi}_{2} + 3H\dot{\varphi}_{2} + V_{2}'(\varphi_{2}) = 0  ,
\end{eqnarray}
where $(\ldots)^{\cdot} \equiv \frac{\partial}{\partial t}$ and
$(\ldots)^{'} \equiv \frac{\partial}{\partial \varphi}$. Following Barrow and
Saich (1993) we assume a simple hypothesis that kinetic and potential energies
of the $\varphi_{1}, \varphi_{2}$ fields are proportional, so
\begin{eqnarray}
\alpha_{1}V_{1}(\varphi_{1}) = \frac{1}{2}\dot{\varphi}_{1}^2  ,\\
\alpha_{2}V_{2}(\varphi_{2}) = \frac{1}{2}\dot{\varphi}_{2}^2  ,
\end{eqnarray}
with $\alpha_{1}, \alpha_{2} =$ const.. If we use the conformal time (2.2)
then we can write down the solutions of (4.4)-(4.5) under the conditions
(4.6)-(4.7) in the following way
\begin{equation}
\varphi_{i, \tau} = \Delta_{i} R^{\left( 1 - \frac{3 \alpha_{i}}{1 +
\alpha_{i}}
 \right)}  ,
\end{equation}
where $\Delta_{i}$ are constants of integration.
The rewritten Friedman equation (2.1) becomes ($j = 1,2,3$)
\begin{equation}
\left( \frac{dR}{d\tau} \right)^2 = C_{j} R^{ - 3 \nu_{j} + 4} + \frac{1}{2}
\left( 1 + \frac{1}{\alpha_{i}} \right) \Delta_{i}^2
R^{\left( 4 - \frac{ 6 \alpha_{i}}{1 + \alpha_{i}} \right)} - k R^2  .
\end{equation}
Comparing (2.1) with (4.9) we realize that in order to exchange
string-like-matter and wall-like-matter for scalar fields we have to take
\cite{D89}
\begin{eqnarray}
\nu_{1} & = & 0 ,  C_{1} = \frac{\Lambda}{3} \nonumber ,\\
\nu_{2} & = & 1 ,  C_{2} = C_{m} \nonumber ,\\
\nu_{3} & = & \frac{4}{3} ,  C_{3} = C_{r} \nonumber ,\\
\alpha_{1} & = & \frac{1}{2} , C_{s} = \frac{3}{2}\Delta_{1}^2 \Rightarrow
\Delta_{1} = \sqrt{\frac{2}{3}\left( k' - k \right)}  ,\\
\alpha_{2} & = & \frac{1}{5} , C_{w} = 3\Delta_{2}^2 \Rightarrow
\Delta_{2} = \sqrt{\frac{\beta\Lambda_{c}^{\frac{1}{2}}}{3}} =
\sqrt{\frac{C_{w}}{3}}    .
\end{eqnarray}
In order to obtain the asymptotic solution (3.9) the exact scalar
fields should be
\begin{equation}
\varphi_{1} = \Delta_{1}\tau + \varphi_{01}
\end{equation}
for $\alpha_{1} = \frac{1}{2}$ and
\begin{equation}
\varphi_{2} = \ln \left[ \varphi_{02} \left| \cosh{\frac{1}{2}\sqrt{\frac{k'}
{2}}\tau} \right| \right]^{ 2 \Delta_{2} \left( 3k'\Lambda_{c}^{\frac{1}{2}}
\right)^{ - \frac{1}{2}}}
\end{equation}
for $\alpha_{2} = \frac{1}{5}$, with $\varphi_{01}, \varphi_{02} =$ const. and
$\Delta_{1}, \Delta_{2}$ given by (4.10)-(4.11). We can thus write down the
required solutions for the scalar fields as
\begin{eqnarray}
\varphi_{1} & = & \sqrt{\frac{2}{3} \left( k' - k \right)}\tau + \varphi_{01}
 ,\\
\varphi_{2} & = & \ln \left[ \varphi_{02} \left|
 \cosh{\frac{1}{2}\sqrt{\frac{k'}
{2}}\tau} \right| \right]^{\frac{4}{9}\sqrt{k'}}  ,
\end{eqnarray}
with the associated potentials
\begin{eqnarray}
V_{1}(\tau) & = & \frac{\Delta_{1}^2}{R^2} = \frac{3}{8} \left( k' - k \right)
k'^2\Lambda_{c}\coth^2{\frac{1}{2}\sqrt{\frac{k'}{2}}\tau}  ,\\
V_{2}(\tau) & = & \frac{5}{2} \frac{\Delta_{2}}{R} = \frac{15}{16\sqrt{2}}
\Lambda_{c}^{\frac{3}{4}} \coth{\frac{1}{2}\sqrt{\frac{k'}{2}}\tau}  ,
\end{eqnarray}
with cosmic time $t(\tau)$ given by (3.10). The potentials as functions
of the scalar fields are (Fig.8)
\begin{eqnarray}
V_{1}(\varphi_{1}) & = & \frac{3}{8} \left( k' - k \right) k'^2 \Lambda_{c}
\coth^2{ \left[ \frac{1}{4} \sqrt{\frac{3k}{k - k'}} \left( \varphi_{1}
- \varphi_{01} \right) \right]}  ,\\
V_{2}(\varphi_{2}) & = & \frac{15}{16\sqrt{2}}k'^2\Lambda_{c}^{\frac{3}{4}}
\coth{ \left[ \cosh^{-1} {\left( \varphi_{02}^{-1} \exp{ \left(
\frac{9\varphi_{2}}{4\sqrt{k'}} \right)} \right)}
\right]}  .
\end{eqnarray}

Of course one could also obtain exact potentials for exact models (3.6) and
(3.11), but they do not oscillate as well. The general solution for
oscillating models is given in terms of elliptic functions by (2.10) ( or
(2.13)
if $\alpha = 0$) and the exact potentials could be calculated using Eq.(4.8).
As another important example we consider the elementary oscillating model given
by
\cite{KAR}.

According to (2.2) the Friedman equation (4.9) in terms of cosmic time
$t(\tau)$ is
\begin{equation}
\left( \frac{dR}{dt} \right)^2 = C_{j} R^{ - 3 \nu_{j} + 2} + \frac{1}{2}
\left( 1 + \frac{1}{\alpha_{i}} \right) \Delta_{i}^2
R^{\left( 2 - \frac{ 6 \alpha_{i}}{1 + \alpha_{i}} \right)} - k  .
\end{equation}
The exact solution given by \cite{KAR} for the case without matter and
radiation is given by (2.26), the values of constants $\alpha_{1}$ and
$\alpha_{2}$ are given by (4.10)-(4.11) and the resulting fields are
\begin{equation}
\varphi_{1}(t) = \varphi_{01} + \sqrt{\frac{2}{3} \left( 1 - \frac{k}{k'}
\right)}
\arctan{\left[ \frac{ C_{w} \tan{\frac{1}{2}t\sqrt{ - \frac{\Lambda}{3}}} + A}
{\sqrt{ - \frac{4}{3} \Lambda k'}} \right]}  ,
\end{equation}
but it reduces to a constant, if there is no strings (i.e. if $k' = k = +1$),
and
\begin{equation}
\varphi_{2}(t) = - \frac{2\Lambda}{3} \sqrt{\frac{C_{w}}{3}}
\int \frac{dt}{\sqrt{C_{w} + A \sin{t \sqrt{ - \frac{\Lambda}{3}}}}}  =
\frac{4}{3} \sqrt{ - \Lambda C_{w}} {\cal P}^{-1} \left[ \tan{\frac{t}{2}
\sqrt{- \frac{\Lambda}{3}}} \right]   ,
\end{equation}
where ${\cal P}^{-1}$ is the inverse function to the Weierstrass elliptic
${\cal P}$ function with invariants given by
\begin{eqnarray}
g_{2} & = & \frac{1}{3} \left( C_{w}^{2} - 4 \Lambda k' \right)   ,\\
g_{3} & = & - \frac{1}{9} C_{w} \left( \frac{C_{w}^{2}}{3} + 4 \Lambda k'
\right)  .
\end{eqnarray}
It is easy to check that (4.21) and (4.22) are oscillatory in t. According to
(4.6), (4.7), (4.10) and (4.11)
\begin{eqnarray}
V_{1}(t) & = & \sqrt{\frac{2}{3} \left( 1 - \frac{k}{k'} \right)}
\frac{C_{w}}{4\sqrt{k'}} \frac{1}{\cos^2{\frac{1}{2} t \sqrt{ -
\frac{\Lambda}{3}}}} \frac{ - \frac{4}{3} \Lambda k'}{\left[ C_{w}
\tan{\frac{1}{2} t \sqrt{ - \frac{\Lambda}{3}}} + A \right]^2 -
\frac{4}{3} \Lambda k'}   ,\\
V_{2}(t) & = & \frac{2\Lambda^2}{3} \sqrt{\frac{C_{w}}{3}}
\frac{1}{C_{w} + A \sin{t \sqrt{ - \frac{\Lambda}{3}}}}   .
\end{eqnarray}
{}From (4.25) and (4.26) we realize that the potentials are oscillatory in t
i.e.
$V_{1}$ oscillates between $B(-4/3\Lambda k')(C_{w} - A)$ and
$B(-4/3\Lambda k')(C_{w} + A)$, where \\$B = 4^{-1}k'^{-
\frac{1}{2}}\sqrt{2/3(1 -
k/k')}$ and $V_{2}$ oscillates between $2\Lambda^{2}/3 \sqrt{C_{w}/3}/(C_{w} -
A)$ and $2\Lambda^{2}/3 \sqrt{C_{w}/3}/(C_{w} + A)$. If we express explicitly
potentials in terms of fields we have
\begin{eqnarray}
V_{1}(\varphi_{1}) & = & \frac{k'^{-2}C_{w}^{-2}}
{32\sqrt{\frac{2}{3} \left( k' - k \right)}}
\left[ \frac{C_{w}^2 + \left[ \sqrt{ - \frac{4}{3} \Lambda k'}
\tan{\sqrt{\frac{3k'}{2 \left(k' - k \right)}}\left( \varphi_{1} - \varphi_{01}
\right)} - A \right]^2}{1 + \tan^2{\sqrt{\frac{3k'}{2 \left(k' - k \right)}}
\left( \varphi_{1} - \varphi_{01}\right)}} \right]^2   \\
V_{2}(\varphi_{2}) & = & \frac{2 \Lambda^{2}}{3} \sqrt{\frac{C_{w}}{3}}
\frac{1}{C_{w} + A \sin{\left\{ 2 \arctan{\left[ {\cal P} \left( \frac{4}{3}
\sqrt{ - \Lambda C_{w}} \varphi_{2} \right) \right]} \right\}}}   .
\end{eqnarray}

The potential $V_{1}(\varphi_{1})$ oscillates between $V_{d} = - \frac{4}{3}
\Lambda k'$ and $V_{u} = 4C_{w}^2 + 4\Lambda k'$. A prove that
$V_{2}(\varphi_{2})$ is also oscillating in $\varphi_{2}$ is more complicated.
The discussion involves the analysis of the discriminant $\Delta = g_{2}^{3} -
g_{3}^{2}$ with $g_{2}$ and $g_{3}$ given by (4.23)-(4.24) and then the
analysis of the solutions similar to that given in Section 2 for the general
oscillatory solution subject to the conditions (2.27). Finally, the result is
that $V_{1}(\varphi_{1})$ and $V_{2}(\varphi_{2})$ in (4.27)-(4.28) do
oscillate in $\varphi_{1}$ and $\varphi_{2}$.

These results together with (4.18) and (4.19) may suggest that at least under
the assumption of proportionality of kinetic and potential energies (4.6)-(4.7)
the monotonic type of the solution for the scale factor $R(t)$ leads to
the monotonic dependence of the potential $V$ on the scalar field $\varphi$ and
the oscillating type of solution for $R(t)$ leads to the oscillating
dependence of the potential $V$ on the scalar field $\varphi$.

\vspace{.6cm}
\section{DISCUSSION}
\vspace{.6cm}

In this paper we discussed a class of nonsingular oscillatory universes
following Kardashev's \cite{KAR} discussion of admitting essential fraction of
stable domain walls as the exotic matter and the negative cosmological
constant. In the context of the dark non-baryonic matter problem both kinds of
fluids can serve as dark matter candidates. The former acts as a source of
repulsive gravity but the latter acts as a source of attraction. The balance
between
these two sources allows the universe to fall into oscillations. In fact, this
is very similar situation to the case of the Einstein Static Universe, where
there is a balance between nonrelativistic positive pressure matter and
repulsive positive cosmological constant. According to our analysis the
Einstein
static models exist as a result of a balance between the exotic matter
and the attractive negative cosmological constant. Unlike the original Einstein
Static Universe these models seem to be mechanically stable as they lie at the
bottom of the asociated mechanical potential (2.7), but their stability should
be examined more carefully.

Since the value of the negative cosmological constant can be reduced to be
very close to zero for oscillatory solutions (cf. Section 2), the most severe
problem of possible compatibility with observations refers to exotic matter,
especially to domain walls \cite{TUR}. Because of that we proposed a
replacement of the exotic matter by scalar fields which may serve as dark
matter candidates as well. We inversed the standard procedure - first we had
the equation of state and then we derived exact potentials \cite{BAR}.
The potentials we studied seem to keep the same properties as the
solutions for the scale factors. It means that monotonic solutions for $R(t)$
(cf.(3.9)) give monotonic potentials (cf.(4.18)-(4.19)) and oscillatory
solutions (cf.(2.26)) give oscillatory potentials (cf.(4.27)-(4.28)).

It is worth emphasizing that the paper is based on many simplifications and
the question is what are the properties of oscillatory solutions in more
physically realistic situations. Firstly, in the discussion of oscillatory
solutions of Section 2 we have neither considered any non-isotropic geometry
nor non-isotropic fluid description for instance for string-like-matter and
wall-like-matter.
Secondly, in the scalar fields interpretation of Section 4 we have assumed
a very simple proportionality relation for the kinetic and potential energies,
the minimal coupling of the fields to gravity and the lack of interaction
between them. Of course all these assumptions might not necessarily apply
and the validity of the presented results should be considered towards these
points.

Especially, in future work one can consider some non-isotropic cosmologies
\cite{FI1,FI2} and non-minimally coupled fields \cite{FU1,FU2}, but these
approaches will certainly be much more difficult to treat analitycally.
Also, one can discuss oscillating quasi-static universes
in the context of the maximum entropy analysis given by Gibbons
\cite{GIB1,GIB2}.
Finally, we should mention one common feature of our results with the results
of Petry \cite{P3} in his conformal gravity theory. Namely, in both theories
nonsingular oscillating solutions appear for the negative cosmological
constant. How far this results could be true for other gravity theories is
a matter for further consideration.

\begin{center}
{\bf Acknowledgments}
\end{center}

The author wishes to thank to John Barrow, Arne Larsen, Jerzy Stelmach and
David Wands for careful reading of the manuscript and helpful suggestions.

\pagebreak

\pagebreak
\begin{center}
{\bf Figure Captions}
\end{center}
Fig.1. The potential of the associated mechanical system (2.9)  for the
Friedman
equation (2.1) with $k' > 0$ and $\beta < \frac{3}{8} k'^2$. The universe can
oscillate in the central well between $T = 0$ and $T = T_{1} = \frac{3}{4}
\left( k' - \sqrt{k'^2 - \frac{8}{3}\beta} \right)$.

Fig.2. The potential of the associated mechanical system (2.9) for the Freidman
equation (2.1) with $k' > 0$ and $\beta = \frac{3}{8} k'^2$. The universe can
oscillate in the central well between $T = 0$ and $T = T_{1} = T_{2} =
\tilde{T}_{2} = \frac{3}{4}k'$.

Fig.3. The potential of the associated mechanical system (2.9)  for the
Friedman
equation (2.1) with $k' > 0$ and $\frac{3}{8} k'^2 < \beta < \frac{1}{2}k'^2$.
The universe can oscillate in the central well between $T = 0$ and $T = T_{2}
= \frac{1}{2} \left( k' + \sqrt{k'^2 - 2\beta} \right)$.

Fig.4. The elementary periodicity cell for oscillatory solutions. The period
$\omega$ is pure real while the period $\omega^{'}$ is pure imaginary.

Fig.5. The exact oscillatory solution in the cases A, B, C from Section 2. This
solution is given in terms of the Weierstrass $\zeta$ functions by (2.21).

Fig.6. The exact recollapsing solution that appears in the cases A, B, C of
Section 2 (formula (2.13)).

Fig.7. The potential of associated mechanical system (2.22) for Kardashev's
case
(no nonrelativistic matter and radiation). Oscillations appear for
$\lambda < 0$ and $k' > 0$.

Fig.8. The exact potentials for the asymptotic solution (3.9).

\end{document}